\documentclass{article}
\pdfoutput=1
\usepackage{PRIMEarxiv}
\usepackage[utf8]{inputenc} 
\usepackage[T1]{fontenc}    
\usepackage{hyperref}       
\usepackage{url}            
\usepackage{booktabs}       
\usepackage{amsfonts}       
\usepackage{nicefrac}       
\usepackage{microtype}      
\usepackage{lipsum}
\usepackage{fancyhdr}       
\usepackage{graphicx}       
\graphicspath{{media/}}     
\usepackage[abs]{overpic}
\usepackage{latexsym}
\usepackage{lineno,hyperref}
\usepackage[english]{babel}

\usepackage{amsmath,amssymb,mathtools}
\usepackage{algorithm}
\usepackage{algorithmic}
\usepackage{amsthm}
\usepackage{multicol}
\usepackage{lineno}
\newtheorem{prop}{Proposition}[section]
\newcommand{\<}{\langle}
\renewcommand{\>}{\rangle}
\newcommand{\R}{\mathbb{R}}
\newcommand{\ds}{\mathbf{d}}
\newcommand{\N}{\mathbb{N}}
\newcommand{\M}{\mathcal{M}}

\newcommand{\CC}{\mathcal{C}}
\newcommand{\LL}{\mathcal{L}}
\newcommand*{\defeq}{\coloneqq}
\newcommand{\smk}{\mathcal{S}_m^k}
\newcommand{\Sm}{\mathcal{S}}
\newcommand{\simk}{\Sigma}

\newcommand{\ga}{\gamma}

\newcommand{\vphi}{\varphi}
\newcommand*{\SO}{\mathrm{SO}}
\newcommand*{\T}{\mathrm{T}}
\newcommand{\Log}{\mathrm{Log}}
\newcommand{\Exp}{\mathrm{Exp}}

\newcommand{\Ver}{\mathrm{Ver}}
\newcommand{\Hor}{\mathrm{Hor}}
\newcommand{\rd}{\mathrm{d}}

\newcommand{\w}{\stackrel{\omega}{\sim}}
\newcommand{\ti}[1]{\tilde{#1}}
\newcommand{\wti}[1]{\widetilde{#1}}

\DeclareMathOperator{\tr}{trace}

\newcommand*{\quark}{\setbox0\hbox{$x$}\hbox to\wd0{\hss$\cdot$\hss}}

\DeclareMathOperator*{\argminA}{arg\,min}
\usepackage{xcolor}
\newcommand{\todo}[1]{\bgroup\color{red}#1\egroup}

\newcommand{\rev}[1]{\leavevmode #1}

\title{A Hierarchical Geodesic Model for\\Longitudinal Analysis on Manifolds\thanks{\textit{This is a preprint version of an article submitted to the Journal of Mathematical Imaging and Vision \textcopyright{Springer}}}}

\author{
Esfandiar Nava-Yazdani, Hans-Christian Hege, Christoph von Tycowicz \\ \\ 
Zuse Institute Berlin, Germany \\
  \texttt{\{navayazdani, hege, vontycowicz\}@zib.de} \\
}

\begin{document}
\maketitle

\begin{abstract}
In many applications, geodesic hierarchical models are adequate for the study of temporal observations.
We employ such a model derived for manifold-valued data to Kendall's shape space.
In particular, instead of the Sasaki metric, we adapt a functional-based metric, which increases the computational efficiency and does not require the implementation of the curvature tensor. We propose the corresponding variational time discretization of geodesics 
and employ the approach for longitudinal analysis of 2D rat skulls shapes as well as 3D shapes derived from an imaging study on osteoarthritis. Particularly, we perform hypothesis test and estimate the mean trends.
\end{abstract}

\section{Introduction}
\label{sec:intro}
Analysis of time-dependent data has become increasingly important for a wide range of applications such as understanding the onset and progression of diseases, physical performance assessment from biomechanical gait data, or facial expression analysis in video sequences.
All these examples can be understood as longitudinal data, where individual instances of a common underlying process are observed at multiple time points.
While common statistical tools like mean-variance analysis and regression allow to study phenomena across individuals or within a single one, longitudinal data exhibits correlations due to repeated measurements that violate the independence assumptions of such cross-sectional methods.
Another issue that warrants attention is missing data that arise, e.g. due to acquisition errors or when subjects drop out of a clinical study.
Proper statistical inference for longitudinal data must therefore account for the within-individual correlation of observations as well as for the sparse and non-uniform sampling.






In this light, hierarchical models and in particular mixed-effects models pose an adequate and very flexible framework for longitudinal data analysis~\cite{locascio2011overview,GFS2016}.
Such approaches deal with the mass of the inherent interrelations by specifying a model in which each subject is assumed to have an own unique functional relation between the dependent variable and time-related predictor(s).
Thus, a parametric spatio-temporal model that optimally fits the data for each given individual is estimated.
Due to random error variation in the dependent variable at each time point for each individual, the fit is generally not perfect.
The coefficients describing these subject-specific models (e.g.\ intercept and slope in the case of straight lines) are in turn assumed to vary randomly in the population.
Corresponding to these random effects, there is a ‘fixed’ effect that is often of primary interest, i.e.\ a single group ‘fixed’ coefficient that indexes the average spatio-temporal model for the entire group.

While there is abundant literature on the mixed-effects framework for scalar and vector-valued measurements, generalizations to other types of data where the domains are structured (such as shapes or graphs) are still at an early stage of research.
However, there is a substantial body of work that demonstrates the advantages of leveraging the structure (or geometry) of the data such as improving the assessment of subject-specific clinical outcomes~\cite{Ambellan2021bScore}, computer-aided diagnosis~\cite{Ambellan2021FCM,seo2016covariant,vonTycowicz2020}, or comparing populations~\cite{MF2012,hong2015group,Hanik2020biinvariant} to name but a few instances.
This provides a strong impetus for the development of generalized hierarchical models that benefit from a compact encoding of constraints and exhibit a superior consistency as compared to their Euclidean counterparts.

A common workaround when facing manifold-valued data is to resort to linear statistical tools by employing a vector space representation---either given explicitly by the discrete representation or derived via a dedicated operation.
For example, in the field of medical image analysis linear mixed-effects models have been applied to vertex coordinates of meshes in order to study shape data~\cite{datar2012mixed,MFK2016}.
In general, however, the quality of the obtained model depends on the validity of the linearity assumption, which is a poor choice for data with a large spread or within regions of high curvature in shape space and, thus, is considered a limiting factor for the ability to represent natural biological variability in populations (see e.g.~\cite{vonTycowicz2018} and the references therein).

To derive coordinate-free, manifold-based formulations Riemannian geometry provides a suitable generalization of straight lines called geodesics that serve as a building block for inference models.
This approach has been explored in multiple works~\cite{MF2012,singh2013hierarchical,hong2015group,muralidharan2017diss} leading to geodesic hierarchical models that encode both subject-specific as well as group trends in terms of geodesics.
There are further extensions that employ a multi-geodesic approach to model developments as function of multiple, possibly categorical covariates~\cite{hong2019hierarchical}, as well as nonparametric formulations~\cite{campbell2018nonparametric} describing group average trends via weighted Fr\'echet means.
In lieu of intrinsic noise distributions that feature tractable computation of the likelihood, model estimation in these approaches is formulated in terms of least-squares criteria instead of maximum likelihood or empirical Bayes estimates.
Nonetheless, under certain conditions the least-squares solutions can be shown to agree with maximum likelihood estimates~\cite{muralidharan2017diss}.

More recently, probabilistic formulations of Riemannian mixed-effects models~\cite{kim2017riemannian,schiratti2015learning,schiratti2017bayesian,bone2020learning,debavelaere2020learning} have been proposed.
These approaches are based on a notion of parallelity that constraints the ‘slopes’ to be fixed for the entire population, whereas subjects in the study may follow different patterns of disease progression.
This assumption can, hence, be a limiting factor diminishing the flexibility and fidelity of these approaches.

In order to gauge differences in geodesic trends many approaches employ a product metric that measures the distance between privileged points on the trajectories as well as the directional deviation (change in ‘slope’). The latter relies on a notion of transport between tangent spaces to spatially align trajectories, e.g.\ parallel transport~\cite{chakraborty2017geometric,kim2017riemannian,hong2019hierarchical} in Riemannian manifolds or co-adjoint transport~\cite{younes2008transport,singh2013hierarchical} in the group of diffeomorphisms.
However, the choice of transport can have a significant impact on the analysis: While the parallel transport depends on the path of transport (an effect called holonomy), the co-adjoint transport is not compliant with the metric.
From a geometrical point of view, more appropriate distances can be derived by considering the space of trajectories itself as a differentiable manifold and equipping it with a Riemannian metric that is in turn consistent with the metric of the data manifold.
State-of-the-art approaches, therefore, identify geodesics with points in the tangent bundle of the data manifold~\cite{MF2012}.
While the Sasaki metric is a natural metric on the tangent bundle, its geodesic computations require time-discrete approximation schemes involving the Riemannian curvature tensor.
This not only incurs high computational costs, but also has a negative effect on numerical stability.

We consider a novel approach that overcomes these shortcomings. To this end, we identify elements of the tangent bundle with vector fields along the geodesic trend.
This provides a notion of a canonical metric that is motivated from a functional view of parameterized curves in the shape space~\cite{srivastava2016functional}. 
Considering the space of the geodesics as a submanifold in the space of shape trajectories, this allows in particular the use of a naturally induced distance. The corresponding shortest path, log map and average geodesic, can be computed by variational time-discretization. 
Remarkably, the underlying energy function allows for fast and simple evaluation increasing computational efficiency.
In particular, the computation of the distance requires neither the computation of the curvature nor the decomposition into horizontal and vertical components.

In this work, we employ the derived metric within the generative hierarchical approach introduced in~\cite{MF2012} that is based on a least-squares theoretic formulation.
In the first stage of this hierarchical model, inner-individual changes are modeled as geodesic trends, which in the second level are considered as disturbances of a population-averaged geodesic trend.
%
%
For the first stage, estimates at the individual level amounts to solving geodesic regression problems~\cite{F2013} for each individual.
These problems can be solved efficiently in terms of first-order Riemannian optimization schemes~\cite{NHST2020}. The involved geodesic computations require geometric quantities such as adjoint Jacobi fields and parallel transport. These important geometric quantities are in general not given as closed form expressions. Efficient approximation schemes have been presented in~\cite{Lorenzi2011,louis2018fanning,NHST2020}.

Using the derived metric for geodesic trends, we obtain a notion of mean, covariance, and Mahalanobis distance.
This allows us to develop a statistical hypothesis test for comparing the group-wise mean trends.
Non-parametric permutation tests are applied to test the significance of the estimated differences in group trends.
We perform this using a manifold-valued Hotelling $T^2$ statistic analogously to~\cite{MF2012} by applying it to the tangent bundle.
As example application we demonstrate the methodology on rat skulls growth and the long term study of incident knee osteoarthritis (OA).

This paper is organized as follows.
In the next section, we describe the geodesic mixed effects model and present an alternate approach for the computation of mean trends as well as the corresponding algorithmic realization. Section ~\ref{sec:ken} provides an overview of Kendall's shape space. Section~\ref{sec:appl} presents the application of our approach to synthetic spherical data for simulation studies, followed by rat calivaria and femur data from an epidemiological longitudinal study dealing on osteoarthritis in Kendall' shape space and a discussion of the numerical results.

\section{Hierarchical Geodesic Model}
\label{sec:geomix}

In this section, we describe the geodesic hierarchical model following least-squares theoretic derivation introduced in~\cite{MF2012}.
In particular, we derive a novel, functional-based Riemannian metric on the space of geodesics, thereby inducing notions of mean geodesics, co-variances in populations of geodesics, and least-squares estimates.

\begin{figure*}[h]
\label{fig:geomix}
\begin{center}
   \includegraphics[width=.8\textwidth]{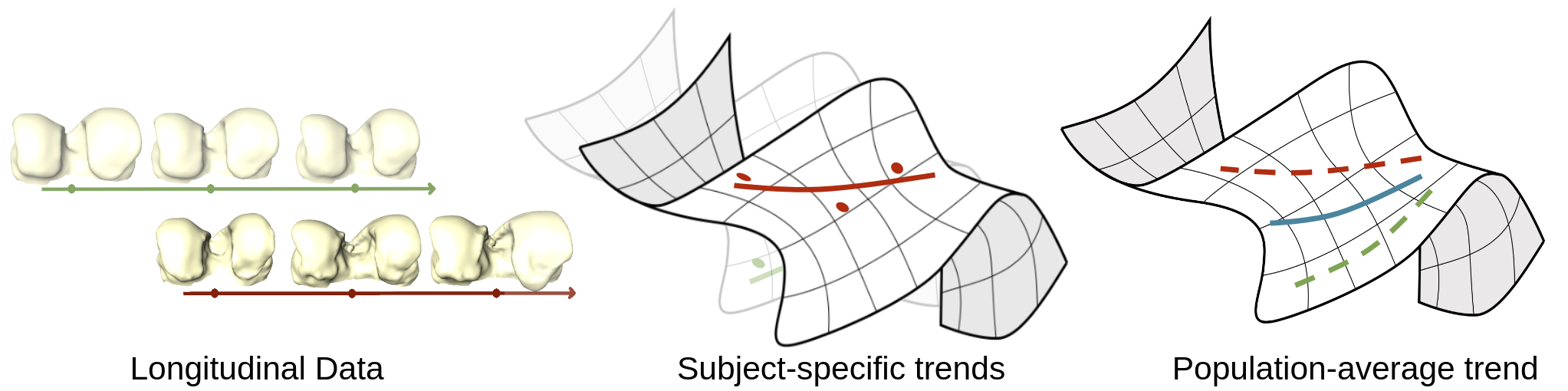}
	\caption{Schematic view of the geodesic mixed effects model: Subject-wise trends represented as best-fitting geodesics and subsequent population-average trend as mean of these geodesics in the shape space.}
\end{center}
\end{figure*}

Let $(M,g)$ be a smooth complete Riemannian manifold with distance function $d$, exponential map $\exp$, its local inverse $\log$ and injectivity radius $r_{inj}$.
Then, the response $Y_i$ of the $i$-th individual is modelled by
\[
Y_i = \exp_{\gamma_i(T_i)}(\epsilon_i),
\]
where $\gamma_i$ denotes the geodesic parameterized with respect to the independent variable $T_i$.
The random variation of observations from the geodesic $\gamma_i$ is modelled by the exponential map and the random variable $\epsilon_i$ that takes values on the tangent space at corresponding points on $\gamma_i$.
The group-level trend is then determined as the mean $\bar\gamma$ geodesic modelled by the exponential map $\exp^{TM}$ of the tangent bundle $TM$
\[
\gamma_i = \exp^{TM}_{\bar\gamma}(\nu_i),
\]
where $\nu_i \in T_{\bar\gamma}TM$ represents the residual geodesic encoding the difference to the group-level geodesic.

To estimate the parameter of the described longitudinal model, we employ a two-step least-squares procedure requiring an appropriate notion of distance in the space of geodesics. 


\subsection{Geodesic Regression}

The first stage of the employed model poses an instance of geodesic regression, which we summarize below.
As such inconsistencies, e.g.\ due to acquisition noise and reconstruction errors, in the individual's observations are minimized.
For an overview and applications we refer to \cite{NHST2020}, \cite{F2013} and \cite{SX2009}.

Consider scalars $t_1 < t_2 < \cdots < t_N$ and distinct points $q_1,\cdots,q_N\in M$. Geodesic regression aims at finding a geodesic curve that best fits the data $q_i$ at $t_i$ in the least-squares sense, i.e., minimizing
\[
\sum_{i=1}^Nd^2(q_i,\ga(t_i))
\]
over $\gamma$ in the space of geodesics. The minimizer is called best-fitting geodesic for data $(t_i,q_i)_{i=1,\cdots,N}$. Applying a linear bijection, we may assume that $0=t_1<t_2<\cdots<t_N=1$.

For a geodesic $\ga$ from $x$ to $y$, $\Phi$ denotes the parametrization of $\ga$ as a path over the unit interval, viz. 
\begin{align}
\label{eq:phi}
\Phi(x,y,t)\defeq\exp_x(t\log_xy),\,t\in I,
\end{align}
where $I\defeq [0,1]$.

Computationally, we employ the above parametrization to determine the corresponding $\Phi(x^\ast,y^\ast,.)$, 
where $(x^\ast,y^\ast) \defeq \argminA F$ and
\[
F(x,y):=\sum_{i=1}^Nd^2(q_i,\Phi(x,y,t_i))
\]
over $M\times M$. The choice $(q_1,q_N)$, serves as a natural initial guess.

\rev{We remark, that one could identify geodesics with their initial points and velocities instead of using endpoints. We employ the latter, since geodesic computations in terms of the function $\Phi$ defined in equation \eqref{eq:phi}, the so-called slerp (spherical linear interpolation), are more efficient.} The predictive power of the regression model can be measured by the coefficient of determination, denoted $R^2$. To compute it, let $F_{min}:=F(x^\ast,y^\ast)$ and denote the minimum of 
\[
G(x)\defeq \sum_{i=1}^N d^2(q_i,x)) 
\] 
by $G_{min}$. Then 
\[
R^2=1 - \frac{F_{min}}{G_{min}}.
\]
We recall that $\frac{1}{N}F_{min}$ and $\frac{1}{N}G_{min}$ are the unexplained and total variance, respectively. Moreover, $R^2\in [0,1]$ and a large value indicates better regression performance (goodness of fit).

Generally, due to absence of an explicit analytic solution, the regression task has to be solved numerically. To this end, we employ a Riemannian trust-regions solver~\cite{manopt} with a Hessian approximation based on finite differences. The main challenge in this regard, is the computation of the gradient of $F$, $\nabla F$. Due to the fact that 
\[
\nabla \rho_y(x)=-2\log_xy,
\]
where $\rho_y(x):=d^2(x,y)$, $\nabla F$ is given by the adjoint of the sum of certain Jacobi fields. For details and application to Kendall's shape space, we refer to \cite{NHST2020}.

\subsection{Tangent bundle and Mean Geodesic}
Geodesic mixed effects models and particularly mean geodesic (group trend) require a notion of distance for the tangent bundle consistent with the 
Riemannian metric of the shape space. In the following, we present a brief introduction to a natural choice for such a distance provided by the Sasaki metric employed in ~\cite{MF2012}. Then, we propose an alternative $L^2$-type approach and induced variational time-discrete geodesics.

Let $\tau$ denote the canonical projection of the tangent bundle $TM$. Suppose that $TM$ is endowed with a Riemannian metric $\rev{\tilde{g}}$. Identifying a geodesic with its endpoints, the mean of geodesics is determined by $\tilde{g}$. A prominent natural choice for $\tilde{g}$ 
is the the Sasaki metric. It is uniquely determined by the following properties (cf. \cite{Sasaki1958}): a) $\tau$ becomes a Riemannian submersion ($\tau$ has maximal rank and $\rd \tau$ preserves lengths of horizontal vectors). b) The restriction of $\tilde{g}$ 
to any tangent space coincides with the Euclidean metric induced by $g$. c) Parallel vector fields along arbitrary curves in $M$ are orthogonal to their fibers, i.e., for any 
curve $\ga$ and parallel vector field $v$ along it, $\dot{v}\perp T_\ga M$.

Let $\eta:=(p,u):I\to TM$ be a smooth curve. $\tau$ being a Riemannian submersion, $T_{\eta}TM$ enjoys an orthogonal decomposition in vertical (viz. kernel of $d_{\eta}\tau$) and 
horizontal subspaces, each of dimension $dim(M)$. Identifying each of them with $T_{p}M$, the Sasaki metric at $\eta$ is induced by the quadratic form $\|v\|^2+\|w\|^2$, 
where $v=p^\prime$ and $w=u^\prime$ \rev{and prime denotes the derivative with respect to the curve parameter $s$}. Denoting the covariant derivative and curvature tensor of $g$ 
by $\nabla$ and $R$, Sasaki geodesics are characterized by 
\begin{align*}
 \nabla_v v &= -R(u,w,v),\\
 \nabla_v w &= 0.
\end{align*}
Algorithms for the computation of the exponential and log map as well as mean geodesic with respect to Sasaki metric, and also an 
application to corpus callosum longitudinal data as trajectories in Kendall's shape space are given in \cite{MF2012}. 
In the case $m=2$ (planar shapes), the shape space can be identified with the complex projective space and the Riemannian curvature tensor 
is explicitly given in terms of the canonical complex structure and the curvature tensor of the pre-shape space. For $m\geq 3$, computation of 
$R$ is more delicate.

We recall that the decomposition of the double tangent bundle into horizontal and vertical subspaces can be expressed in terms of Jacobi fields as follows (cf.\ ~\cite{K1977}). The map \rev{$H(s,t):= \exp_{p(s)} (tu(s))$} is a family of geodesics. Therefore, $J_s(t):=H^\prime (s,t)$ is a family of Jacobi fields. The horizontal and vertical components of $\eta^\prime$ read $v=J(0)$ and $w=\dot{J}(0)$.

Next, we propose an alternative approach to employ a metric on the tangent bundle. Fix $s\in I$ and let $\ga_s:I\to M$ be the geodesic emerging from $p(s)$ with initial velocity $u(s)=\dot{\ga_s}(0)$. 
Let $\xi_s$ be a vector field along $\ga_s$. Then by $\|\xi_s\|_{L^2}^2=\int_Ig(\xi_s (t),\xi_s(t))\,dt$ a quadratic form at $(p(s),u(s))$ is given, which defines a distance, denoted by $\delta$ for the space of geodesics and in turn, on the tangent bundle $TM$. Hence, from an extrinsic point of view, $\delta$ is induced by the standard metric of the Hilbert manifold $L^2(I,M)$.

Now, let $H=H(s,t)$ be a family of paths $I\to M$ ($H(s,\cdot)$ not necessarily geodesics) with $\alpha:=H(0,\cdot)$ and $\beta:=H(1,\cdot)$ geodesics. The energy of $H$ induced by the above quadratic form reads 
\[
E(H)=\int_0^1\int_0^1g(H^\prime(s,t),H^\prime(s,t))\:dt\:ds
\] 
($\xi_s=H^\prime(s,.)$). Let $H^*$ denote the minimizer of $E$ restricted to paths through geodesics, i.e. $H(s,.)$ geodesic for all $s\in I$. 
Fix $n\in\N$. Next, we construct time-discrete paths $(H^*_i)_{i=0,\cdots,n}$ (sequence of geodesics) to approximate $H^\ast$, which in turn, identifying each geodesic with its initial point and velocity, provides a discrete path in the tangent bundle approximating $(H^*(0),\dot{H^*}(0))$. We remark that the energy functional $E$ achieves its minimum in $L^2(I,M)$, i.e. over all paths connecting $\alpha$ to $\beta$, if $H(.,t)$ is a geodesic for all $t$. It follows from a slight modification of \cite[Theorem 3.2]{srivastava2016functional}, which we present in section \ref{sec:geoapp} for the reader's convenience. Now, let $s_i:=\frac{i}{n}$. We may write
\[
E(H)=\sum_{i=0}^{n-1}\int_{s_i}^{s_{i+1}}\|H^\prime(s,.)\|_{L^2}^2\,ds
\]
and discretizing $E$ intrinsically, i.e.\ replacing $H^\prime(s,.)$ by the finite difference $n\log_{H_i} H_{i+1}$ with $H_i(.):=H(s_i,.)$, we arrive at
\begin{align*}
E_n(H)=n\sum_{i=0}^{n-1}\delta^2(H_i,H_{i+1}).
\end{align*}
Identifying each geodesic $H_i$ with its endpoints $x_i=H_i(0)$ and $y_i=H_i(1)$, we have to minimize the explicit expression 
\begin{equation}
\label{log}
E_n((x_i,y_i)_{i=0,\cdots,n})
=n\sum_{i=0}^{n-1}\int_0^1 d^2(\Phi (x_i,y_i,t),\Phi (x_{i+1},y_{i+1},t))\: dt.
\end{equation}
Now, to ensure that any two points determine a unique geodesic, we suppose that $\alpha$ and $\beta$ are close enough in the sense that their images lie in a convex normal neighbourhood (for instance a geodesic ball with radius  $\frac{r_{inj}}{2}$) and let $(x^\ast,y^\ast)$ be the minimizer of $E_n$
over $(x_i,y_i)_i\in M^n\times M^n$ with fixed endpoints $x_0=\alpha(0),y_0=\alpha(1)$, $x_{n}=\beta(0),y_{n}=\beta(1)$. A natural choice for 
the initial values $x^0$ and $y^0$ is given by the equidistant partition $x^0_i=\Phi(x_0,x_{n},s_i)$, 
$y^0_i=\Phi(y_0,y_{n},s_i)$. Then, the desired discrete shortest path reads  
\begin{equation}
\label{path}
    H^*=\left(\Phi(x^\ast_0,y^\ast_0,.),\cdots,\Phi(x^*_n,y^*_n,.)\right).
\end{equation}  
We remark that the discrete path length is given by 
\begin{align*}
\LL_n((x_i,y_i)_i)&=\sum_{i=0}^{n-1}\delta(H_i,H_{i+1})\\
&=\sum_{i=0}^{n-1}\sqrt{\int_0^1 d^2(\Phi (x_i,y_i,t),\Phi (x_{i+1},y_{i+1},t))\:dt}.
\end{align*}
Furthermore, for any discrete path $H$ one has the apriori estimation  
\[
\delta^2(H_0,H_n)\leq n\sum_{i=0}^{n-1}\delta^2(H_i,H_{i+1}),
\]
with equality if and only if all $H_i$ equidistantly lie along a smooth shortest path from $H_0$ to $H_n$. 

Note that $H_0=\alpha$, $H_n=\beta$ and the right hand side of the above inequality is the discrete energy $E_n(H)$, explicitly given by \eqref{log}. 
Indeed, one verifies that, if $M$ is the Euclidean space, then the minimizer coincides with the initial value and in the above estimation equality holds. 
Moreover, it coincides with the discrete Sasaki geodesic and converges to the Sasaki geodesic given by \begin{align*}
    p(s)&=(1-s)\alpha (0)+s\beta (0),\\ u(s)&=s\dot{\alpha}(0)+(1-s)\dot{\beta}(0).
\end{align*} 
Next, we aim at constructing the mean of $N$ geodesics $\ga_j:I\to M$. To ensure well-defindeness of the approach, we suppose that they are close enough, 
in the sense that their images lie in a neighbourhood $U$ within $\frac{1}{2}min \{r_{inj},\frac{\pi}{\sqrt{\Delta}}\}$, where $\Delta$ denotes an upper bound 
for the sectional curvature in $U$ and $\frac{\pi}{\sqrt{\Delta}}$ is interpreted as $+\infty$ if $\Delta\leq 0$ (e.g. for Hadamard manifolds).
Now, the induced mean of $\ga_1,\cdots,\ga_N$ is the geodesic $\ga$ with initial- and endpoints $x$ and $y$, minimizing over $M^2$
\begin{equation}
\label{mean}
 G_n(x,y)=\sum_{j=1}^N \min_{x_i^j, y_i^j \in M}\,E_n((x^j_i,y^j_i)_{i=0,\cdots,n})\quad  \text{s.t.}\,\,x^j_0=x,\,y^j_0=y,\,x^j_n=\ga_j(0),\,y^j_n=\ga_j(1).
\end{equation}

A natural choice for the initial value is given by the point-wise mean of $(\ga_j(0),\ga_j(1))$, $j=1,\cdots,N$. In the sequel, we call $\delta$ the functional-based $L^2$-metric. 

Note that computations of the log map and mean with respect to $\delta$ neither involve the curvature tensor nor decomposition in horizontal and vertical components.

\subsection{Algorithmic Realization}
In this section, we describe the algorithmic realization of the computation of the variational time-discrete shortest path and the mean geodesic with respect to the proposed $L^2$-metric, denoted by $\delta$.

Discrete shortest paths are determined as minimizers of the discrete path energy $E_n$ given in~\eqref{log}.
Before we turn to the general case, let us first consider solutions of $E_2$, i.e.\ discrete 2-geodesics in $TM$.
Assuming endpoints to be fixed, i.e.\ seeking solutions of the boundary value problem, the search space for solutions of $E_2$ reduces to a single $M$-geodesic.
Indeed, given a quadrature rule for $t$-discretization, the estimation of 2-geodesics numerically coincides with (weighted) geodesic regression. Explicitly, we minimize over $x,y\in M$
\begin{align*}
    E_2((x_-,y_-),(x,y),(x_+,y_+))
    =\sum_{i=0}^k\omega_id^2(\Phi (x,y,t_i),\Phi (x_{-},y_{-},t_i))
    +d^2(\Phi (x,y,t_i),\Phi (x_{+},y_{+},t_i)))),
\end{align*}
where discrete values $t_i$ and weights $\omega_i$ are determined by the chosen quadrature rule.
This analogue allows us to reuse algorithms and implementations for geodesic regression that are already required during the first stage of our hierarchical model.
Moreover, as the solutions to $E_2$ are also midpoints of the corresponding 2-geodesic, we can interpret them as discrete averages.
This allows to derive an optimization procedure for general $E_n$ by adopting a discrete path shortening flow from the realm of polygonal path processing.
In particular, we employ an iterative averaging approach detailed in Algorithm~\ref{alg:short} that consecutively updates each of the inner nodes by replacing it with the average of its neighbours as determined by $regression$, i.e.\ estimation of a 2-geodesic ($n=2$).
One verifies that this algorithms converges, as each averaging step weakly decreases the discrete path energy $E_n$.

\begin{algorithm}
\caption{Discrete Shortest Path}
\label{alg:short}
\begin{algorithmic}
\REQUIRE Initial and final geodesics $\alpha$ and $\beta$
\ENSURE Shortest path $(x_i,y_i)_i \in M^n \times M^n$ from $\alpha$ to $\beta$
\STATE $(x_i,y_i)_i \leftarrow (\Phi(\alpha(0),\beta(0),i/n), \Phi(\alpha(1),\beta(1),i/n))_{i=0,\ldots,n}$

\REPEAT
\FOR{$i = 1,\cdots,n-1$}
\STATE $(x_i,y_i) \leftarrow regression((x_{i-1},y_{i-1}), (x_{i+1},y_{i+1}))$
\ENDFOR
\UNTIL{convergence}
\end{algorithmic}
\end{algorithm}

Computation of the discrete mean geodesic $\bar{\gamma}$ requires minimization of Eq.~\eqref{mean} yielding a simultaneous optimization over $N$ discrete paths $(x_i^j,y_i^j)_{i=0,\ldots,n}$ connecting the input geodesics $\gamma_j$ and $\bar{\gamma}$.
While interior nodes of each path have to form 2-geodesics with their neighbours, the coupling at the center is described as minimizer of $G_1$, where input geodesics are the first interior nodes $(x^j_1,y^j_1)$ along the discrete paths.
Again, the optimization of $G_1$ w.r.t.\ the center $\bar{\gamma}$ can be considered as an instance of geodesic regression following a similar argument as for $E_2$.
This motivates an alternating optimization procedure that---in every iteration---first relaxes the estimate for $\bar{\gamma}$ and subsequently updates the interior nodes of the paths each time solving a geodesic regression problem.
To improve efficiency, we further employ a cascadic approach in time by solving the sequence of related problems $(\min_\gamma G_k)_{k=1,\ldots,n}$ using the $k$-th solution to initialize the subsequent problem.

\section{Kendall's Shape Space}
\label{sec:ken}
In the following we present a brief overview of Kendall's shape space and its tangent bundle as well as main quantities which will be employed for geodesic analysis and statistics. Note that our approach is not limited to the Kendall's shape space and can be employed to any Riemannian manifold.

For a comprehensive introduction to Kendall's shape space and details on the subjects of this section, we refer to \cite{KBC1999} and \cite{NHST2020}. For the relevant tools from Riemannian geometry, we refer to \cite{GHL2004} and \cite{Oneill66}.

Let $M(m,k)$ denote the space of real $m\times k$ matrices endowed with its canonical scalar product given by $\<x,y\>=\tr(xy^t)$, 
and $\| \quark \|$ the induced Frobenius norm. 
We call the set of $k$-ad of landmarks in $\R^m$ after removing translations and scaling the pre-shape space and identify it with \[\smk \defeq \{x\in M(m,k):\: \sum_{i=1}^kx_i=0,\:\|x\|=1\}\] endowed with the standard spherical distance given by $d(x,y)=\arccos (\< x,y\>)$. 

A shape is a pre-shape with rotations removed. More precisely, the left action of $\SO_m$ on $\smk$ given by $(R,x)\mapsto Rx$ defines an equivalence relation given by $x\sim y$ if and only if $y=Rx$ for some $R\in \SO_m$.
\emph{Kendall's shape space} is defined as $\simk=\smk/\mathord{\sim}$. Now, denoting the canonical projection of $\sim$ by $\pi$, the induced distance between any two shapes $\pi(x)$ and $\pi (y)$ is the Procrustes distance  given by
\begin{equation*}
	d(x,y)= \min_{R\in \SO_m} d(x,Ry)=\arccos \sum_{i=1}^m\lambda_i,
\end{equation*}
where $\lambda_1\geq \cdots\geq \lambda_{m-1}\geq  |\lambda_m|$ denote the pseudo-singular values of $yx^t$. 
Note that for simplicity of notation, we have identified shapes and their representing pre-shapes in the definition of $d_{\Sigma}$. 
We call $x,y\in \smk$ \emph{well positioned} and write $x\w y$ if and only if $yx^t$ is symmetric and $\ds(x,y)=d(x,y)$.
For each $x,y\in \smk$, there exists an optimal rotation $R\in \SO_m$ such that $x\w Ry$. The diameter of $\sigma$ is $\pi/2$. 

Furthermore, the horizontal and vertical spaces at $x\in \smk$ read
\begin{align*}
	\Hor_x &=\{u\in M(m,k-1):\:ux^t=xu^t\text{ and } \<x,u\>=0\},\\
	\Ver_x &=\{Ax:\:A+A^t=0\}.
\end{align*}
A smooth curve is called horizontal if and only if its tangent field is horizontal.
Geodesics in the shape space are equivalence classes of horizontal geodesics.
For $x\w y$ the geodesic from $x$ to $y$ given by
\begin{equation*}
	\Phi(x,y,t)=\frac{\sin((1-t)\vphi)}{\sin\vphi}x+\frac{\sin(t\vphi)}{\sin\vphi}y
\end{equation*}
with $\vphi=\arccos(\<x,y\>),\:0\leq t\leq 1$, is horizontal. 
Hence $\Phi$ realizes the shortest path from $\pi(x)$ to $\pi(y)$. Now, let $\Log$ and $\Exp$ denote the log and exponential map of the pre-shape sphere, i.e., 
\[
\Log_xy=\vphi\frac{y-\langle x,y\rangle x}{\| y-\langle x,y\rangle x \|},
\]
and
\[
\Exp_xv=\cos (\vphi)\cdot x+\frac{\sin (\vphi)}{\vphi}\cdot v,
\]
with $\|v\|=\vphi,\,\langle x,v\rangle =0$. 
Then the Riemannian exponential map of the shape space, denoted by $\wti{\exp}$, satisfies \[\pi(\Exp_x u)=\wti{\exp}_{\pi(x)}(\rd_x\pi(u))=\wti{\exp}_{\pi(x)}(\rd_x\pi(u^h)),\]
where $u^h$ stands for the horizontal component of $u\in T_x\smk$. Furthermore, the log map of the shape space is represented at $x$ by $(\Log_x)^h$.

We recall, that pre-shapes with rank $\geq m-1$, denoted by $\Sm$, constitute an open and dense subset of $\smk$ and the restriction of the quotient map $\pi$ to $\Sm$ is a Riemannian submersion with respect to the metric induced by the ambient Euclidean space. Moreover, for pre-shapes in $\Sm$, the optimal rotation is unique if and only if $\lambda_{m-1}+\lambda_m\neq 0$. Denoting the covariant derivatives in the pre-shape and shape space by $\nabla$ respectively \ $\ti{\nabla}$, for horizontal vector fields $X$ and $Y$, we have
\[
\nabla_XY=(\nabla_XY)^h+\frac{1}{2}[X,Y]^v.
\]
Here the superscript $v$ denotes the vertical component and $[.,.]$ the Lie bracket. Moreover, $(\nabla_XY)^h$ equals the horizontal lift of $\ti{\nabla}_{\rd\pi X}\rd\pi Y$. In particular, 
\[
	\rd \pi(\nabla_X Y)=(\ti{\nabla}_{\rd \pi X} \rd \pi Y) \circ \pi.
\]
We recall that denoting the Euclidean derivative of a vector field $W$ along a path $\ga$ in the pre-shape sphere (i.e.  $\|\ga\|=1$) by $\dot{W}$, we have \[\nabla_{\dot{\ga}}W=\dot{W}-\<\dot{W},\ga\>\ga.\]

Key quantities of the shape space geometry such as parallel transport, Jacobi fields and Fr{\'e}chet mean can be computed by horizontal lifting 
to $\Sm$ (and extension to $\smk$). We refer the reader to \cite{NHST2020} for corresponding results.

\section{Experiments}
\label{sec:appl}
In this section, we provide experimental evaluations based on the proposed geodesic hierarchical model for synthetic as well as publicly available, longitudinal data.
First, we perform a simulation study investigating the efficacy of the proposed method. Second, we provide a comparison of mean trends w.r.t.\ the Sasaki and our proposed metric, apply the approaches to planar shapes describing rat skulls and compare the results. Subsequently, we apply the derived scheme to the analysis of group differences in longitudinal femur shapes of subjects with incident and developing osteoarthritis (OA) versus normal controls. In particular, we estimate group-wise trends and perform a Hotelling $T^2$ test to identify significant group differences.
For two-dimensional visualization, we apply tangent PCA at the Fr{\'e}chet mean of shapes representing the observed data. Time discrete computations are performed based on 2-geodesics---employing finer discretizations have been found to provide no further improvements for the datasets under study.

\subsection{Synthetic Spherical Data}
We performed simulation studies with responses on the 2-dimensional unit sphere $S^2$. The objective of these studies is to illustrate the application of the proposed approach and to verify its efficacy. To this end, we fixed a geodesic $\mu$ with initial- and endpoints $\mu_0$ and $\mu_1$, and employed two distributions of spherical geodesics with mean $\mu$, to generate random geodesics. We call the distributions $M^2$ and Sasaki. Let $\sigma=\frac{\pi}{12}$. The $M^2$ distribution uses the parametrization of geodesics over $S^2\times S^2$. Therein a sample geodesic $\ga$ is given by initial- and endpoints $\ga_0$ and $\ga_1$ drawn from a normal distribution with standard deviation $\sigma$ centered at $\mu_0$ and $\mu_1$, respectively. For the Sasaki distribution, we randomly generate vectors $v$ and $w$ in $T_{\mu_0}S^2$ from a normal distribution with standard deviation $\sigma$ and mean zero. We gain  $\ga_0$ and $\dot{\ga }(0)$ by applying the Sasaki exponential map at $(\mu_0,u)$ to $(v,w)$, where $u=\dot{\mu}(0)$. We used $\frac{1}{\pi}\max (d( \mu_0,\ga_0),d(\mu_1,\ga_1))$ with $d(x,y)=\arccos (\langle x,y\rangle )$, to measure the distance between $\mu$ and $\ga$ and thus, quantify the accuracy of our estimations. The scaling factor $\frac{1}{\pi}$ reflects the fact that the error is bounded above by $\pi$, the diameter of the sphere.

\begin{figure}[h]
    \centering
    \includegraphics[width=.35\textwidth]{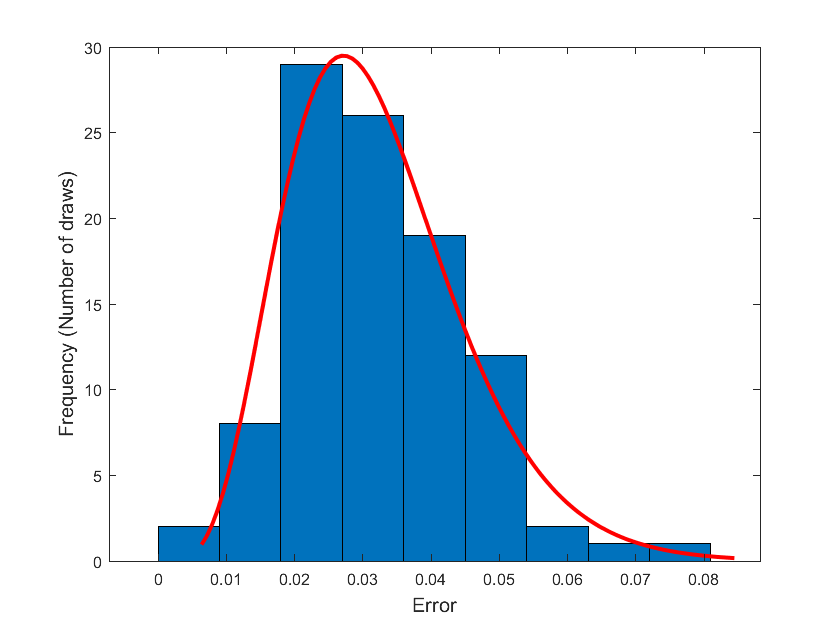}
    \includegraphics[width=.35\textwidth]{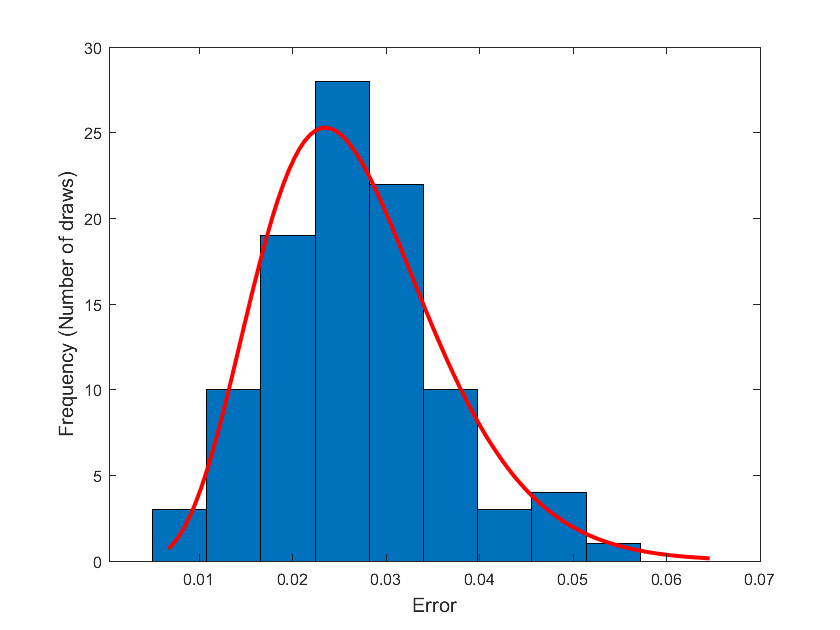}
    \includegraphics[width=.35\textwidth]{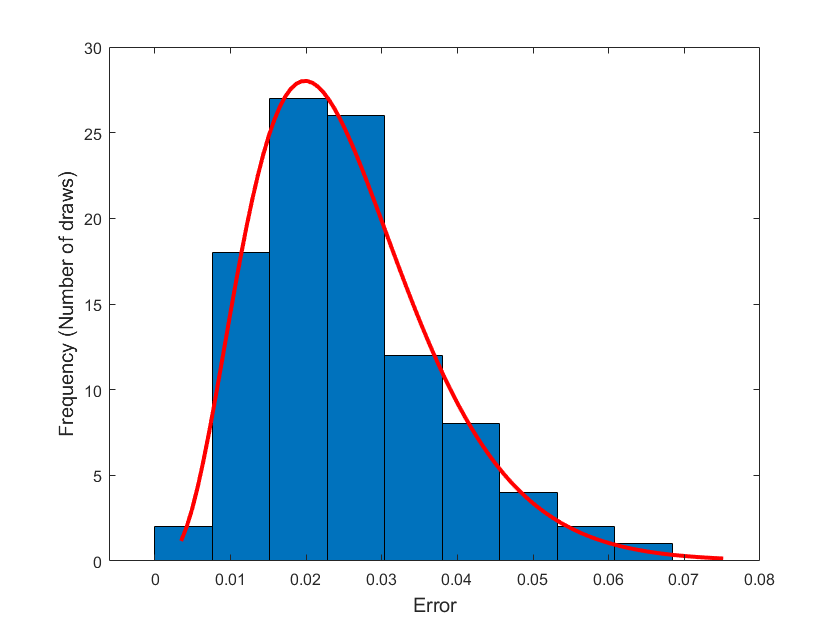}
    \includegraphics[width=.35\textwidth]{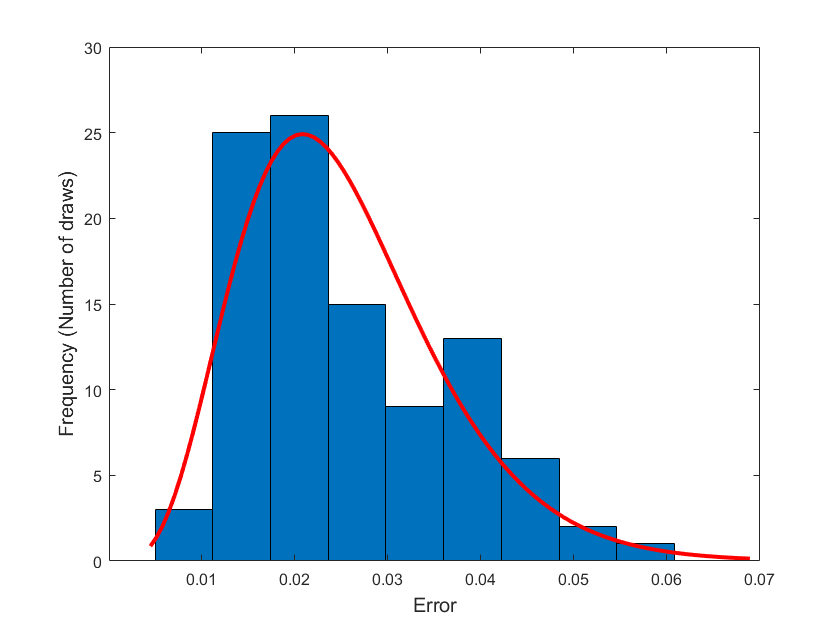}
	\caption{Frequency distribution of error for  estimated means of 25 random geodesics for 100 draws from the $M^2$ (top) and Sasaki distribution (bottom) for Sasaki (left) and proposed metric (right).
	}
\label{fig:simul_repeat}
\end{figure}

In the first study, we repeatedly (100 times) estimated mean geodesics from 25 randomly generated geodesics. 
The resulting bias (average error) and median (95\% confidence intervals) from the $M^2$ distribution amount to 0.0269 and 0.0259 (0.0250–0.0288), and 0.0323 and 0.0304 (0.0297–0.0348) for the proposed and Sasaki metric, respectively. For both metrics, the Sasaki distribution yields almost identical values, viz. 0.0253 and 0.0234, with slightly different confidence intervals (0.0231–0.0275) and 0.0234 (0.0230–0.0277). Fig.~\ref{fig:simul_repeat} shows the frequency distribution for this study with a fitted density function from a beta distribution. This choice of density function is motivated by the boundedness of the error function and the opening that its probability is unknown.  

\begin{figure}[h]
    \centering
    \includegraphics[width=.45\textwidth]{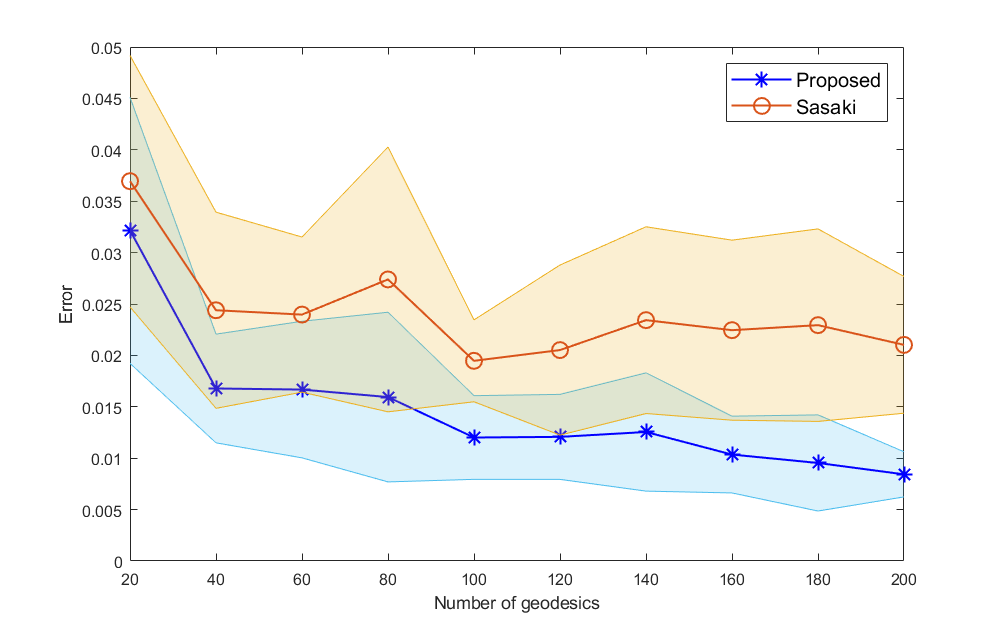}
    \includegraphics[width=.45\textwidth]{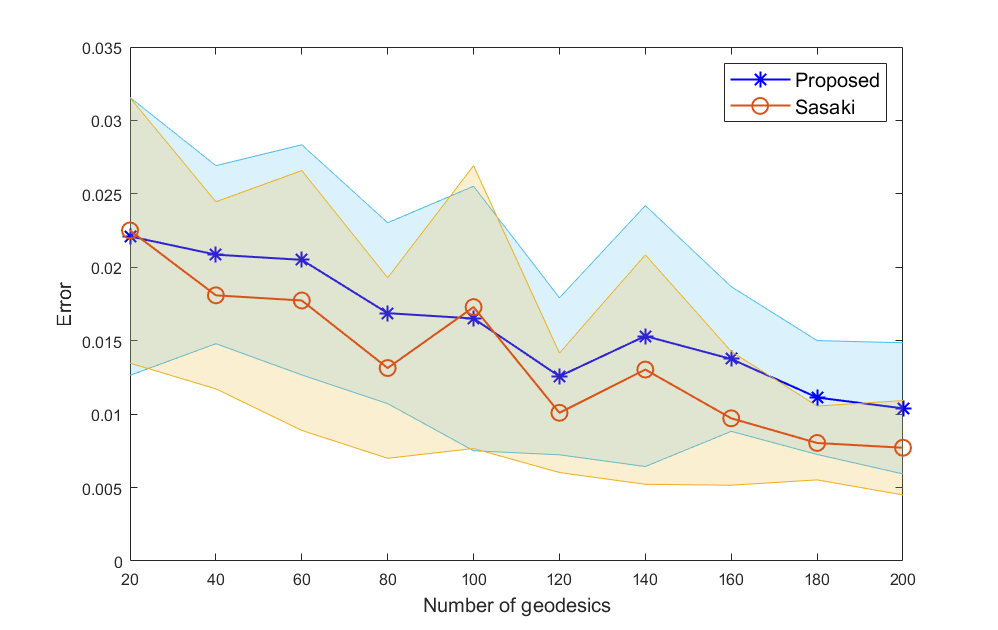}
	\caption{Average error of mean estimates versus increasing number of geodesics for 10 draws from $M^2$ (left) and Sasaki distribution (right). Shaded regions indicate standard deviations.
	}
\label{fig:simul_increase}
\end{figure}

In the second study, in each experiment, we computed the error of the estimated mean w.r.t. both approaches, and increased the number of geodesics. We repeated the experiment 10 times and  observed that for both approaches the error  clearly exhibits an overall decreasing trend (as expected). For both distributions the deviation between the proposed approach and the Sasaki one is reasonable. Fig.~\ref{fig:simul_increase} shows a plot of the average errors $\pm$ standard deviations against the number of geodesics. However, the overall trend for the Sasaki metric has slightly more fluctuations, while the error for proposed metric exhibits a relatively more uniform trend. The former approach uses numerical second order covariant derivatives and spherical parallel transport causing more computational sensibility, while the latter one is based on quadrature. This seems to be a reasonable explanation for the mentioned difference. Regarding computation time, there was no significant difference between the approaches.

We remark that one could use another distance function instead of the one induced by the canonical metric on $S^2\times S^2$. However, we observed that the other natural choices, Sasaki distance or maximum of $d(\mu (t),\ga (t))$ over $t\in I$, only slightly change the error values, but qualitatively have no remarkable effect on the considered essential statistical characteristics, i.e., frequency distribution or average trends.

\begin{figure}[h]
\begin{center}
    \includegraphics[width=.3\textwidth]{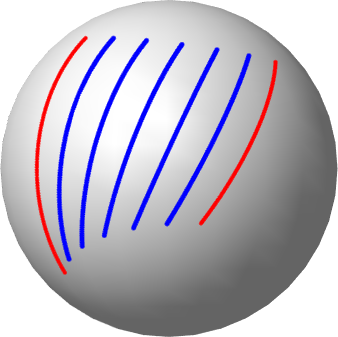}\hspace{.03\textwidth}
   \includegraphics[width=.3\textwidth]{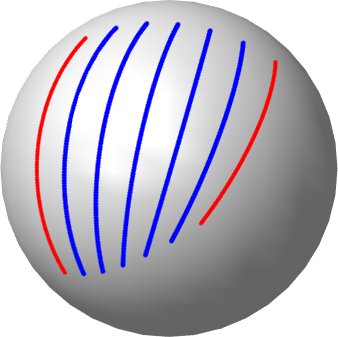}
	\caption{Minimal geodesic in the tangent bundle identified as shortest path connecting two geodesics (red) with respect to Sasaki (left) and functional-based $L^2$-metric (right).}
\label{fig:sas_str_geo}
\end{center}
\end{figure}

\begin{figure}[h]
    \centering
    \includegraphics[width=.3\textwidth]{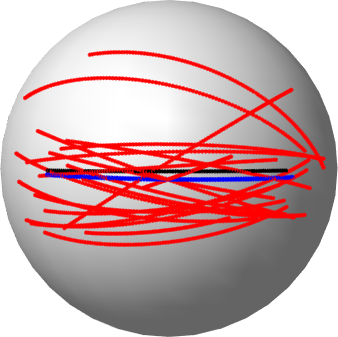}\hspace{.03\textwidth}
    \includegraphics[width=.3\textwidth]{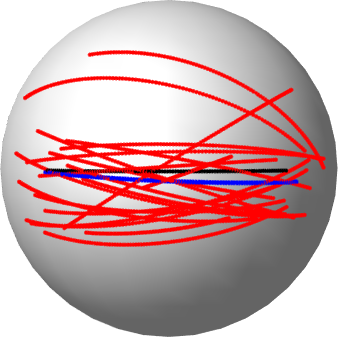}
	\caption{Red geodesics are generated by randomly perturbing the endpoints of the black one. Their mean geodesic with respect to Sasaki (left) and 
	functional-based $L^2$-metric (right) are blue. Both approaches provide an adequate approximation of the true  mean.
	}
\label{fig:sas_str_mean}
\end{figure}

A qualitative comparison between the shortest path and mean induced by the Sasaki and the proposed metric is shown in Fig.~\ref{fig:sas_str_geo} and~\ref{fig:sas_str_mean}, respectively. For none of the shortest paths footpoint curves constitute geodesics. However, the functional-based one is closer to
 the more intuitive shortest path given by the simple point-wise construction $H(s,t)=\Phi(\alpha(t),\beta(t),s)$.

\subsection{Mean Trend for Rats Calvaria Data}
As the first, open-access\footnote{\url{https://life.bio.sunysb.edu/morph/data/datasets}} dataset we use Vilmann's rat calvaria (skulls excluding the lower jaw) that have been obtained from X-ray images and were also studied in \cite[pp. 408-414]{bookstein1997morphometric}.
It consists of 8 landmarks in 2 dimensions for 18 individuals observed at ages of 7, 14, 21, 30, 40, 60, 90, 150 days.
\begin{figure}[h]
    \begin{center}
		\includegraphics[width=.55\textwidth]{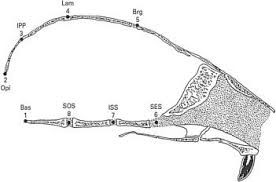}
    \caption{Landmark configuration of rat braincase (laterally viewed and the skull is facing right) with bottom of the figure corresponding to the basicranium (source: \cite{bookstein1997morphometric}).}
	\label{fig:rat}
	\end{center}
\end{figure}
Our computations for this data yield a coefficient of determination of $R^2=0.794$, in accordance with \cite{H2014}.
The 2D Kendall's shape space is isometric to the complex projective space with its standard (Fubini-Study) metric, hence a symmetric space. Therefore, decomposition in vertical and horizontal parts, parallel transport and the curvature tensor can be represented by closed form expressions (cf. section \ref{sec:planarapp}).
Hence, application of our approach to the rat skulls dataset, provides a suitable opportunity for the comparison of the Sasaki and the proposed mean geodesic.

\begin{figure}[h]
    \begin{center}
    \includegraphics[width=.3\textwidth]{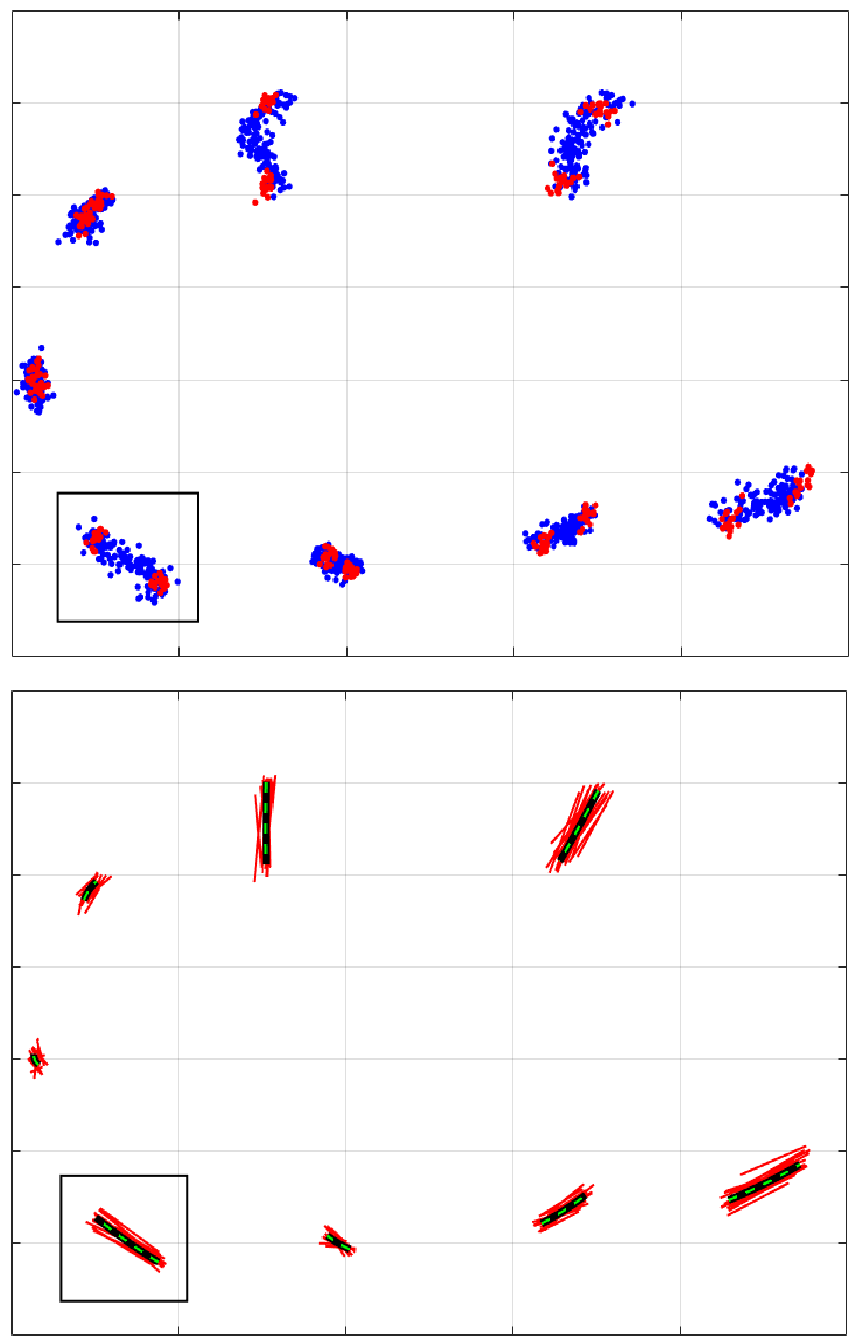}
    \vspace{2ex}
        \includegraphics[width=.3\textwidth]{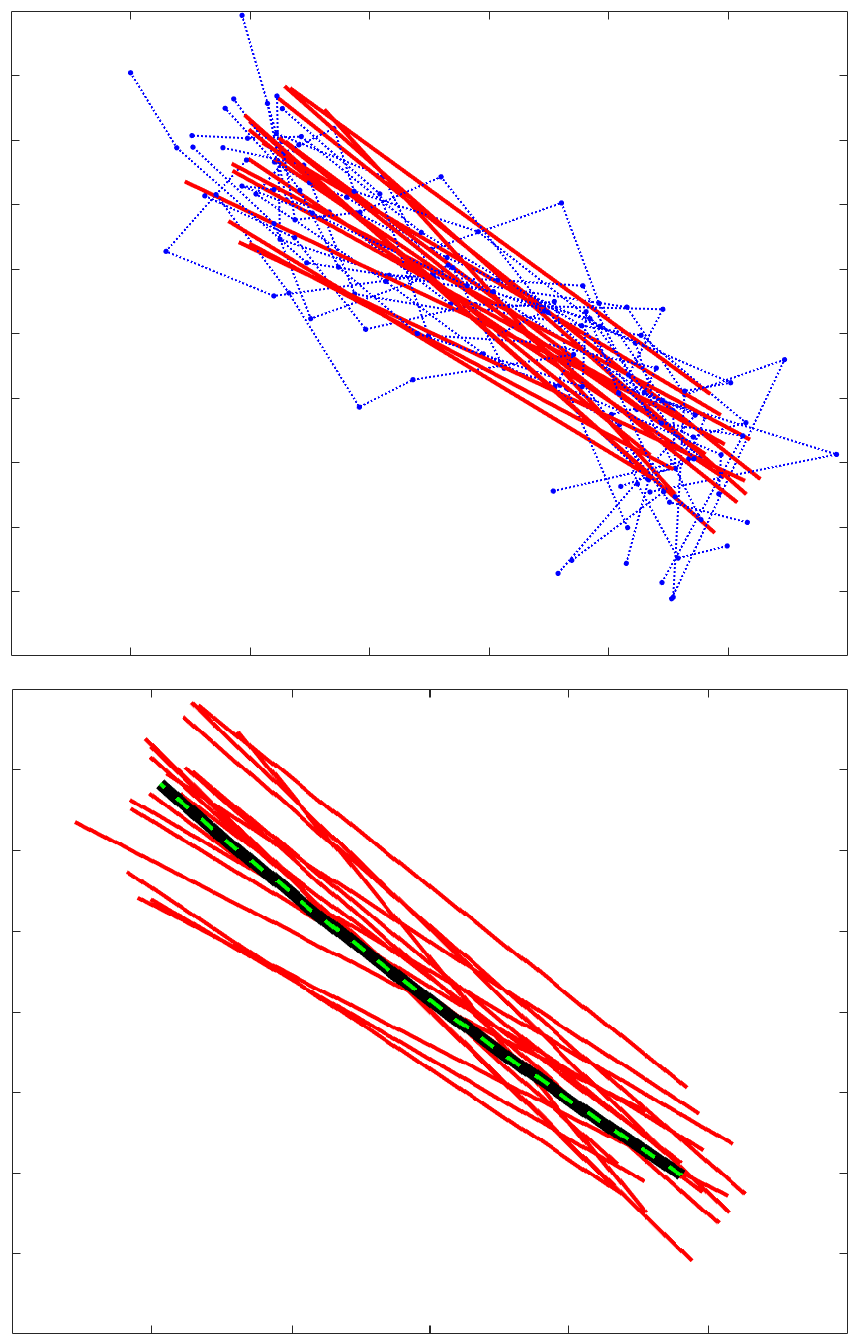}
	\caption{Top:Input shapes (blue) and endpoints of fitted geodesics (red) with zoomed view of input shape trajectories and fitted geodesics for landmark 1. Bottom: Fitted geodesics, Sasaki (black) and proposed (green) mean trend with zoomed view for landmark 1.}
\label{fig:rat_sasstr}
\end{center}
\end{figure}
\begin{figure}[h]
    \begin{center}
    \includegraphics[width=.5\textwidth]{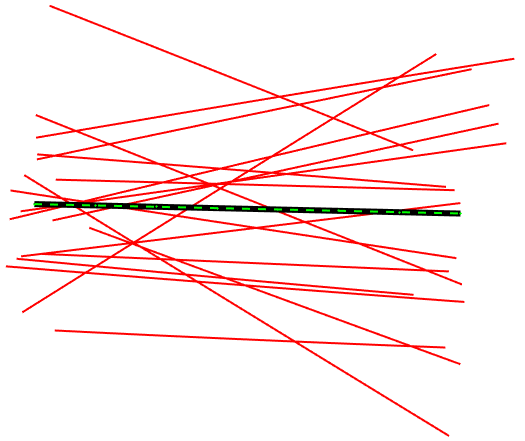}
	\caption{Two-dimensional visualization of fitted geodesics (red) with their Sasaki (black) and proposed (green) mean trend (visually indistinguishable).}
\label{fig:rat_pca}
    \end{center}
\end{figure}

Fig.~\ref{fig:rat_sasstr} visualizes landmark-wise, shape representations of the input and output corresponding to the first stage (top) and second stage (bottom) of the geodesic mixed-effects model applied to rats calvaria data. Fig.~\ref{fig:rat_pca} shows the result of tangent PCA to the fitted geodesics and their mean providing the average overall trend. The deviation between the two mean geodesics induced by the Sasaki and our proposed metric just amounts to $10^{-4}$. Therefore they are not distinguishable n the plots.
\subsection{Group-wise Trends for Femoral Data and Hypothesis Tests}
Our second dataset is derived from the Osteoarthritis Initiative (OAI), which is a longitudinal study of knee osteoarthritis comprising (among others) clinical evaluation data and radiological images from 4,796 men and women of age 45-79. The data are available for public access at \url{http://www.oai.ucsf.edu/}.

From the OAI database, we determined three groups of shapes trajectories: HH (healthy, i.e.\ no OA), 
HD (healthy to diseased, i.e.\ onset and progression to severe OA), 
and DD (diseased, i.e.\ OA at baseline) according to the Kellgren--Lawrence 
score~\cite{kellgren1957KLscore} of grade 0 for all visits, an increase of at least 3 grades over the course of the study, and grade 3 or 4 for all visits, respectively.
We extracted surfaces of the distal femora from the respective 3D weDESS MR images (0.37$\times$0.37 mm matrix, 0.7 mm slice thickness) using a state-of-the-art automatic segmentation approach \cite{Ambellan2018segmentation}.
For each group, we collected 22 trajectories (all available data for group DD minus a record that exhibited inconsistencies, and the same number for groups HD and HH, randomly selected), 
each of which comprises shapes of all acquired MR images, 
i.e.\ at baseline, the 1-, 2-, 3-, 4- 6- and 8- year visits.
In a supervised post-process, the quality of segmentations as well as the correspondence of the resulting meshes (8,988 vertices) were ensured.

\rev{Main goals in this application are the following. First, to understand how anatomical structures in the femur change over time, during aging and growth processes, and more critically during disease progression. Second, to compare and test how temporal changes in the anatomy of different groups significantly differ. }We represented \rev{t}he data in Kendall's shape space and applied the geodesic regression approach described above to the femoral trajectories.

We computed the 
$R^2$-values amounting to medians (95 confidence intervals) of 0.40 (0.33–0.46), 0.55 (0.48–0.63), and 0.51 (0.40–0.72) for group HH, DD, and HD, respectively. For details and computational aspects, we refer to \cite{NHST2020}. 
Note that geodesic representation provides a less cluttered visualization of the trajectory population
making it easier to identify trends within as well as across groups.
\begin{figure}[h]
\begin{center}
    \includegraphics[width=.5\textwidth]{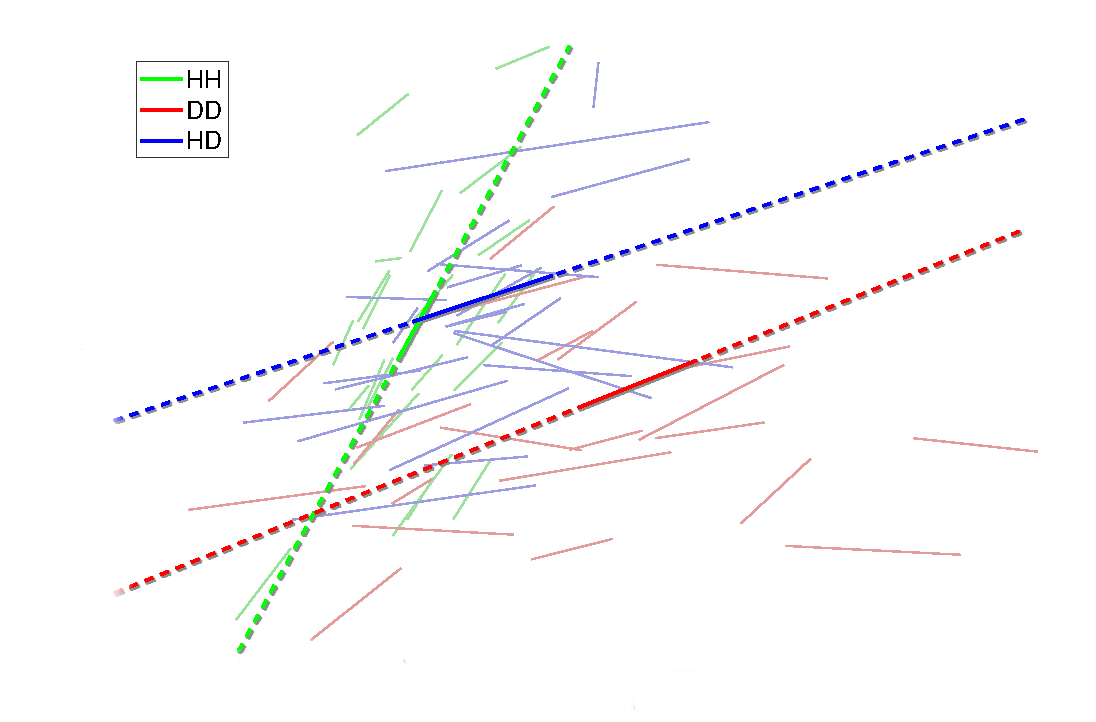}
	\caption{Two-dimensional visualization for group-wise trends estimated as averages of fitted geodesics.
	}
\label{fig:mixed_effects}
\end{center}
\end{figure}

For the statistical testing of group differences, we employ the manifold-valued Hotelling $T^2$ test described in \cite{MF2012}. 
For the convenience of the reader, the formulae used therein are presented below. 
Let $x=\{x_1,\cdots,x_{n_1}\}$ and $y=\{y_1,\cdots,y_{n_2}\}$ two samples with 
corresponding Fr{\'e}chet means $\bar{x}$ and $\bar{y}$, $v_x=\log_{\bar{x}}\bar{y}$, $v_y=\log_{\bar{y}}\bar{x}$. Then the individual group covariances are given by 
\begin{align*}
 W_x &=\frac{1}{n_1}\sum_{i=1}^{n_1}(\log_{\bar{x}}x_i)(\log_{\bar{x}}x_i)^t \\
 W_y &=\frac{1}{n_2} \sum_{i=1}^{n_2}(\log_{\bar{y}}y_i)(\log_{\bar{y}}y_i)^t 
\end{align*}
and the sample $T^2$ statistic reads
\[
 t^2=\frac{1}{2}( v_x^tW^{-1}_x v_x + v_y^tW^{-1}_y v_y ). 
\]
For the estimations of the log map and mean, we employed equations \eqref{log} and \eqref{mean}. 
We found $t^2$-values $0.0012$, $0.000703$ and $0.000591$ for HH vs. DD, DD vs. HD and HH vs. HD with corresponding $p$-values $0$, $0.011$ and $0.033$. 
For the computation of the statistical significance, i.e. $p$-values, we randomly permuted group memberships of the subject-specific geodesic trends  $1,000$ times. 
The results reveal clear differences between the group-wise average geodesics demonstrating the descriptiveness of the proposed approach.
In particular, the results confirm the qualitative differences in group-average trends depicted in the low-dimensional visualization in Fig.~\ref{fig:mixed_effects}.
\section{Conclusion}
We presented a modification of the geodesic hierarchical model introduced in \cite{MF2012} by employing 
a discrete geodesic for the tangent bundle of the shape space instead of Sasaki geodesics. Our approach does not involve 
the Riemannian curvature tensor and allows for simple and efficient approximation. Furthermore, we presented numerical examples, using synthetic random spherical data as well as 2D shapes representing longitudinal data for rat skulls, demonstrating the accuracy of our approach. 

Moreover, we estimated average geodesics and group trends for the example application of femoral longitudinal 3D data using Kendall's shape space, and employed a manifold-valued Hotelling $T^2$ test, which confirmed that the model well distinguishes the groups.

There are several potential directions for future work.
First, it would be interesting to derive efficient expressions for the estimation of Sasaki geodesics for 3D shape trends and to compare the result with our approach.
Furthermore, we would like to extend our model to account for errors-in-covariates, as well as higher-dimensional parameters, which would allow to take further effects into account providing more insight on more complex phenomena.
In the end, further investigation and understanding of distributions on manifolds is needed to obtain probabilistic, yet computationally tractable generalizations of mixed-effects models.

\section*{Acknowledgments}
This work was supported by the Bundesministerium fuer Bildung und Forschung (BMBF) through BIFOLD - The Berlin Institute for the Foundations of Learning and Data (ref. 01IS18025A and ref 01IS18037A) as well as the Deutsche Forschungsgemeinschaft (DFG, German Research Foundation) under Germany´s Excellence Strategy – MATH+ : The Berlin Mathematics Research Center, EXC-2046/1 – project ID: 390685689.
Furthermore, we are grateful for the open-access OAI dataset of the Osteoarthritis Initiative\footnote{
Osteoarthritis Initiative is a public-private partnership comprised of five contracts
(N01-AR-2-2258; N01-AR-2-2259; N01-AR-2-2260; N01-AR-2-2261; N01-AR-2-2262) 
funded by the National Institutes of Health, a branch of the Department of Health and 
Human Services, and conducted by the OAI Study Investigators. Private funding partners 
include Merck Research Laboratories; Novartis Pharmaceuticals Corporation, 
GlaxoSmithKline; and Pfizer, Inc. Private sector funding for the OAI is managed by the 
Foundation for the National Institutes of Health. This manuscript was prepared using an OAI 
public use data set and does not necessarily reflect the opinions or views of the OAI 
investigators, the NIH, or the private funding partners.}.

This article is an extended version of our approach~\cite{NHT2019} first introduced at the MICCAI 2019 workshop on Mathematical Foundations of Computational Anatomy.
Furthermore, note that the proposed hierarchical model has been generalized to higher-order trends in our concurrent work~\cite{HanikHegevonTycowicz2022}. An implementation of the latter (including our geodesic approach as special case) is available as open source 
in~\cite{Morphomatics}.
\section{Appendix}
\subsection{Geodesics in the Space of Curves}
\label{sec:geoapp}
$\M:=L^2(I,M)$ is a manifold and its tangent space at $\ga\in \M$ is given by
\begin{align*}
T_\ga \M=\{\xi:I\to TM\quad | \quad \xi (t)\in T_{\ga (t)}M\mbox{ for all }t\in I \mbox{and }\int_Ig_{\ga (t)}(\xi (t),\xi (t))\,dt<\infty \}.
\end{align*}
Moreover, $\M$ endowed with the inner product
\begin{align*}
\xi_1,\xi_2\in T_\ga\M \mapsto \int_Ig_{\ga (t)}(\xi_1(t),\xi_2(t))\,dt
\end{align*}
is a Riemannian Hilbert manifold (cf. \cite{Michor1980ManifoldsOD}). The following result is borrowed form \cite{srivastava2016functional}, to which we refer for details and example applications.
\begin{prop}
Let $I\ni s\mapsto H(s,t)$ with $t\in I$, be a path in $\M$ and $H_t(s):=H(s,t)$. Suppose that for all $t\in I$, $H_t$ is a geodesic in $M$. Then $H$ is a geodesic in $\M$.   
\begin{proof}
We recall that a path on a Riemannian manifold is a geodesic if and only if the gradient of the energy $E$ with respect to the path is zero. Thus, it suffices to show that the path $H$ is a critical point of the energy $E$. Now, let $C:I^2\times ]-\epsilon,\epsilon[$ be an arbitrary variation of $H$ in $\M$. Thus, $C(s,t,0)=H_t(s)$, $C(0,t,\nu)=H_t(0)$ and $C(1,t,\nu)=H_t(1)$ for all $\nu\in ]-\epsilon,\epsilon[$ and  $s,t\in I$. Now, fix $\nu\in ]-\epsilon,\epsilon[$. Denoting
\[
f(s,t,\nu)=g_{C(s,t,\nu)}(\dot{C}(s,t,\nu),\dot{C}(s,t,\nu)),
\]
we have
\begin{align*}
E(C)&=\int_0^1\int_0^1 f(s,t,\nu)\,dt\, ds\\
&=\int_0^1\int_0^1 f(s,t,\nu)\,ds\, dt,
\end{align*}
where dot stands for differentiation with respect to $t$. Differentiating with repect to $\nu$ at $\nu=0$, we conlude
\begin{align*}
\frac{d}{d\nu}|_{\nu =0}E(C(.,.,\nu )) &=\int_0^1 \frac{d}{d\nu}|_{\nu =0}\left( \int_0^1 f(s,t,\nu)\,ds \right) dt\\
&=\int_0^1 \left(\int_0^1 \frac{d}{d\nu}|_{\nu =0}  f(s,t,\nu)\,ds\right) dt.
\end{align*}
The above expression is zero, since $s\mapsto C(s,t,0)=H_t(s)$ is a geodesic for every $t$, thus 
$\frac{d}{d\nu}|_{\nu =0}  f(s,t,\nu)$ vanishes identically. \end{proof}
\end{prop}
\subsection{Geometry of Planar Shapes}
\label{sec:planarapp}
For the convenient of reader, we summarize the main mathematical results for planar shapes ($m=2$), which (identifying the shape space with the complex projective space) are well known. They also follow immediately from corresponding results presented in \cite{NHST2020}. 

We denote $\CC=\begin{pmatrix}0 & -1\\ 1 & 0\end{pmatrix}$ and remark that $\CC$ is the involution, defining the almost complex structure and the symmetry of the shape space corresponding to multiplication with the imaginary unit in the representation as complex projective space. In particular, $Ver_x =\R\, \CC$.

Now, let $x\in \Sm$ and $w\in \T_x\Sm$. Then the vertical component of $w$ reads $Ax$, where \[A=wx^t-xw^t=|w|\sin(\alpha)\, \CC\] and $\alpha$ denotes the angle between $w$ and its horizontal component $w-Ax$.

Let $\ga :[0,\tau]\to \Sm$ be a smooth horizontal path with initial velocity $v$, $u$ a horizontal vector at $x \defeq \ga(0)$ and $W$ a vector field along $\ga$ with $W(0)=u$. Let \[C=uv^t-vu^t,\] 
and denote the unit matrix of size $m$ by $Id$. If $\ga$ is a unit-speed geodesic, then the parallel transport of $u$ is given by
		\[
			W=U+(\<u,v\>Id+C)(\dot{\ga}-v),
		\]
		where $U$ denotes the Euclidean parallel extension of $u$ along $\ga$, i.e., $U(t)=u$ f.a.\ $t$. If $y = \ga (\varphi)$ with $\varphi=d (x,y)$, then
		the parallel transport $W_y$ of $u$ along $\ga$ to $y$ reads
		\[
			W_y = U- 2\frac{\<u,y\>Id+C\sin (\varphi)}{\|x+y\|^2}(x+y).
		\]
We recall that $W$ is a representative of the  parallel extension in the shape space (by horizontal lifting) and not simply the horizontal component of the spherical parallel extension.

For horizontal vector fields $X,Y$ and $Z$, the Riemannian curvature tensor of the shape space is given by
\begin{align*}
 R(X,Y)Z &=\langle Y,Z \rangle X-\langle X,Z \rangle Y +\langle \CC Y,Z \rangle \CC X\\ 
\quad &-\langle \CC X,Z \rangle \CC Y 
-2 \langle \CC X,Y \rangle \CC Z.
\end{align*}
In particular, the sectional curvature at a plane spanned by $X$ and $Y$ reads
\[
1+\frac{3\langle X,\CC Y\rangle^2}{\|X\|^2\|Y\|^2-\langle X,Y\rangle^2}.
\]
\bibliographystyle{unsrt}
\bibliography{geomixed_arxiv}
\end{document}